\documentclass[%
 reprint,
superscriptaddress,
 amsmath,amssymb,
 aps,
 prl,
]{revtex4-2}
\pdfoutput=1
\usepackage{ulem}
\usepackage{hyperref}
\usepackage{graphicx}
\usepackage{amsmath}
\usepackage{color}
\usepackage{physics}
\usepackage{dsfont}
\usepackage{comment}
\usepackage{booktabs}
\usepackage{pifont}
\usepackage[T1]{fontenc}
\usepackage{makecell}
\usepackage{pdfpages} % include pdfs
\usepackage{pgffor} % for loops
\usepackage{bm}
\usepackage{braket}

\makeatletter
\AtBeginDocument{\let\LS@rot\@undefined}
\makeatother

\definecolor{limegreen}{rgb}{0.2, 0.8, 0.2}
\definecolor{orange}{rgb}{1.0, 0.5, 0.0}

\definecolor{emerald}{rgb}{0.31, 0.78, 0.47}
\definecolor{blue(ncs)}{rgb}{0.0, 0.53, 0.74}

\hypersetup{
     colorlinks=true,
     linkcolor=magenta,
     filecolor=blue,
     citecolor=emerald,      
     urlcolor =blue(ncs),
}

\begin{document}
\title{Protection of Unconventional Superconductivity from Disorder}

\author{Sofie Castro Holb\ae k}
\affiliation{Department of Physics, University of Zurich, Winterthurerstrasse 190, 8057 Zurich, Switzerland}

\author{Morten H. Christensen}
\affiliation{Niels Bohr Institute, University of Copenhagen, DK-2100 Copenhagen, Denmark} 

\author{Andreas Kreisel}
\affiliation{Niels Bohr Institute, University of Copenhagen, DK-2100 Copenhagen, Denmark} 

\author{Brian M. Andersen}
\affiliation{Niels Bohr Institute, University of Copenhagen, DK-2100 Copenhagen, Denmark}

\date{\today}

%\vskip 1cm

\begin{abstract}
Unconventional superconductivity is a desirable state of matter due to its potential for high transition temperatures $T_{\mathrm{c}}$ and associated favorable superconducting properties. However, the sign-changing nature of the order parameter of unconventional superconductors renders their condensates fragile to disorder, an inevitability in real materials. We uncover the generic properties of electronic band structures and associated Bloch weights able to support robust unconventional superconductivity. We demonstrate this property in several case studies of the kagome and Lieb lattices, showing how unconventional superconductors exhibit unusually weak $T_{\mathrm{c}}$ suppression by disorder, despite featuring fully compensated sign-changing order parameters. We contrast these results with those for unconventional superconductivity on the square and honeycomb lattices, which are unable to protect the condensates from disorder. Finally, we discuss material candidates for which this effect may be realized.
\end{abstract}

\maketitle

\paragraph{Introduction.} A hallmark of unconventional superconductivity is a sign-changing order parameter $\Delta(\bm{k})$~\cite{sigrist:1991,scalapino:2012}. In many cases, this leads to nodes in the gap and the existence of low-energy quasiparticles, crucial for the low-temperature properties of the superconducting condensate. Exceptional cases include some iron-based materials identified as $s^{\pm}$ pairing, where the Fermi surface remains fully gapped, despite the sign-changing nature of the order parameter~\cite{hirschfeld:2011}.
Even for such cases however, the unconventional superconducting states are known to be fragile to disorder, meaning that the critical transition temperature $T_{\mathrm{c}}$ drops rapidly with increasing scattering rate~\cite{balatsky:2006,mackenzie:1998}. A strong $T_{\mathrm{c}}$ suppression is often defined in comparison to the Abrikosov-Gor'kov (AG) theory of the pair-breaking effect of magnetic impurities on $T_{\mathrm{c}}$ of conventional ($s$-wave) superconductivity~\cite{abrikosov:1960}. The sign-changing nature of the order parameter also allows in-gap bound states, often useful for narrowing down the detailed gap structure of unconventional superconductors. This has been discussed in view of cuprates, heavy-fermion materials, Sr$_{2}$RuO$_{4}$, and iron-based superconductors, and has played a pivotal role in the fundamental understanding of unconventional superconductivity~\cite{smith:1985,pan:2000,hudson:2001,rullier:2003,balatsky:2006,fischer:2007,tsai:2009,chi:2016,zhou:2013,yin:2015,kreisel:2015,hirschfeld:2016,kreisel:2020,chen:2023,ranna:2025}. By contrast, a weak $T_{\mathrm{c}}$ suppression and the absence of in-gap bound states are commonly interpreted as evidence against unconventional superconductivity~\cite{anderson:1959}.

Recently, we reported an exceptional case strongly deviating from the expected disorder response of unconventional superconductors: compensated pairing states on the kagome lattice, that is, pairing states where $\Delta(\bm{k})$ sums to zero over the Brillouin zone (BZ) \textit{by symmetry}, are surprisingly robust to disorder, exhibiting weak $T_{\mathrm{c}}$ suppression and an absence of in-gap bound states~\cite{holbaek:2023}, such that unconventional superconductivity appears conventional from the perspective of disorder. In Ref.~\cite{holbaek:2023}, the origin of this effect was traced to the properties of the Fermi surface Bloch states. The existence of disorder-robust unconventional superconductivity raises the tantalizing question of whether there exist other materials, or perhaps engineered systems, with high $T_{\mathrm{c}}$ supported by this mechanism?

\begin{figure*}[t]
    \centering
    \includegraphics[width=\linewidth]{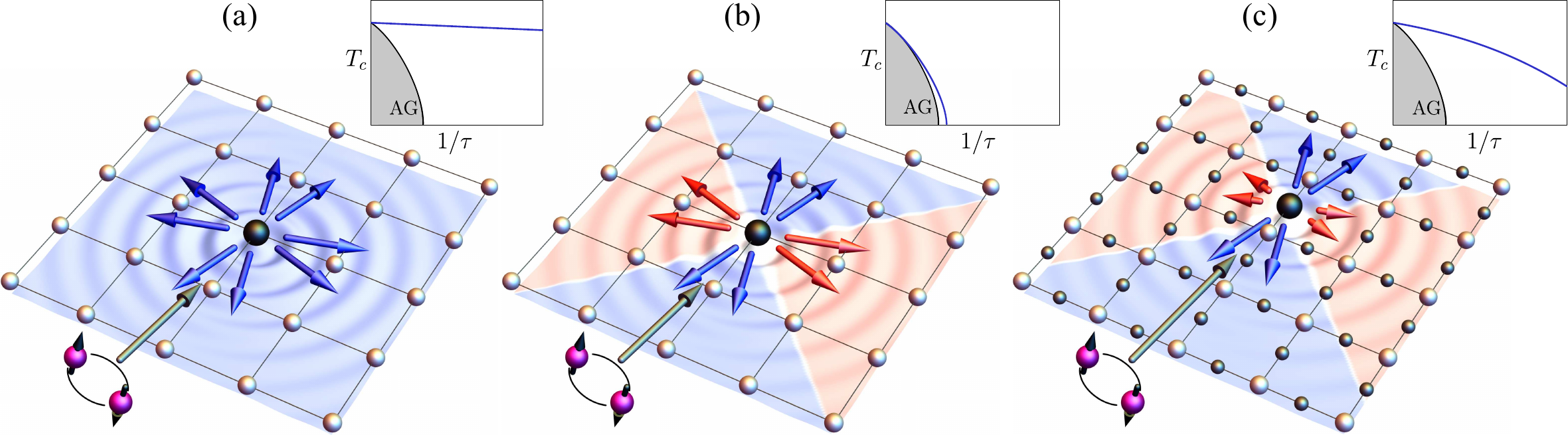}\caption{\label{fig:schematic} \textbf{Schematic illustration of $T_{\mathrm{c}}$ suppression by disorder.} Cooper pair electrons, in purple, scatter off an impurity, in black, with the resulting $T_{\mathrm{c}}$ suppression indicated by the blue line in the insets. The nature of the superconducting condensate is illustrated by the underlying blue and orange regions. In (a), a conventional, non-sign-changing order parameter on the square lattice is shown. Here, the non-compensated nature of the gap results in robust superconductivity. In contrast, (b) shows a compensated order parameter on the square lattice leading to fragile superconductivity. A compensated order on the Lieb lattice is shown in (c). Here, the anisotropy of the Bloch weights reduces scattering along sign-changing directions making the superconductivity more robust to disorder, reminiscent of the conventional case.}
\end{figure*}

In this Letter, we perform a general study of the robustness of unconventional pairing to disorder. We identify the guiding principle for when compensated pairing states feature weak $T_{\mathrm{c}}$ suppression, and illustrate this through several examples including $d$-wave superconductivity on the kagome and Lieb lattices. Broadly speaking, we find that non-trivial superconducting order parameters that transform trivially under the site-symmetry group of the site hosting the impurity atom are robust to disorder. We focus on quasi-two-dimensional systems since unconventional superconductivity is most often found in such crystal structures~\cite{scalapino:2012} and consider point-like disorder. Generated by vacancies or substitutional impurities, the impurity Hamiltonian couples directly to the degrees of freedom that are active in the superconducting pairing.
In other words, the disorder is of the harmful type that typically leads to AG behavior. Figure~\ref{fig:schematic} illustrates the basic differences in impurity effects between conventional and unconventional superconductors. In the conventional case seen in Fig.~\ref{fig:schematic}(a), scattering is non-pair-breaking due to the non-sign-changing nature of the order parameter. In the unconventional case shown in Fig.~\ref{fig:schematic}(b), point-like disorder scatters equally to states with same and opposite sign of the gap function, leading to strong pair-breaking of the condensate. Finally, Fig.~\ref{fig:schematic}(c) illustrates the scenario presented here, where disorder mainly scatters to same-sign directions despite the underlying pair state being fully compensated. An important ingredient for the latter scenario is the presence of symmetry-enforced regions of vanishing sublattice weight of the relevant Bloch states.

\begin{figure}[t]
\centering
\includegraphics[width=\linewidth]{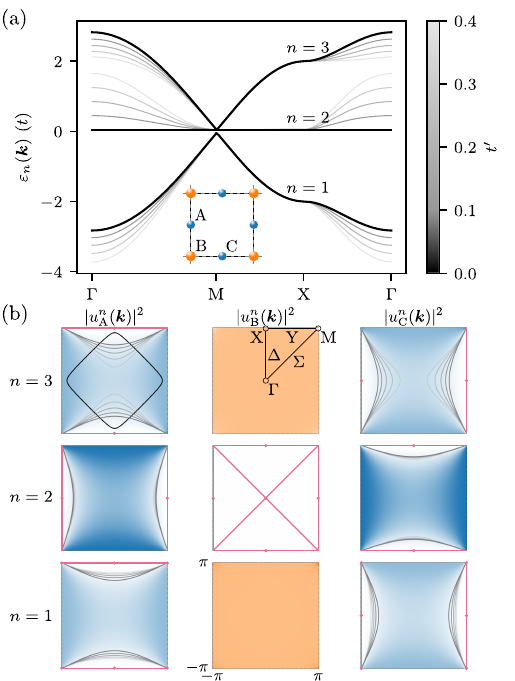}
\caption{\label{fig:Figure2} \textbf{Electronic structure of the Lieb lattice} (a) Energy bands of the Lieb lattice tight-binding model with nearest-neighbor hopping $t = 1$ and varying next-nearest-neighbor hopping $t'$, plotted along $\Gamma-\mathrm{M}-\mathrm{X}-\Gamma$.
The inset defines the three sites of the Lieb lattice. (b) The sublattice weight $|u_{\alpha}^{n}(\bm{k})|^{2}$ for each sublattice A, B, and C, and each band plotted in the BZ. The gray lines illustrate $|u_{\alpha}^{n}(\bm{k})|^{2} = 0.1$ contours for $t' = 0$ (opaque) to $t' = 0.4$ (faint).
The symmetry-enforced nodes of the Bloch weights at high-symmetry points are shown in pink, and the Fermi surface for $\mu = 2.025$ is shown in black.}
\end{figure}

\paragraph{Symmetry-enforced constraints on Bloch weights.}
To demonstrate how such regions emerge, we start from a generic Hamiltonian
\begin{equation}
    \mathcal{H} = \sum_{\substack{\bm{k}\sigma \\ \alpha\beta}} h_{\alpha\beta}(\bm{k})c^{\dagger}_{\alpha\bm{k}\sigma}c^{\phantom{\dagger}}_{\beta\bm{k}\sigma}\,,
\end{equation}
where $c^{\dagger}_{\alpha\bm{k}\sigma}$ ($c_{\alpha\bm{k}\sigma}$) creates (annihilates) an electron with momentum $\bm{k}$, spin $\sigma$, on sublattice $\alpha$.
The Hamiltonian is diagonalized through a unitary transformation
\begin{equation}
    c_{\alpha\bm{k}\sigma} = \sum_n u_{\alpha}^{n}(\bm{k}) c_{n\bm{k}\sigma}\,,
\end{equation}
with $n$ labeling the energy eigenstate. The Bloch eigenstates $u_{\alpha}^{n}(\bm{k})$ encode how the contribution from sublattice $\alpha$ to the band $n$ is distributed across the BZ. Depending on the symmetries of the sublattice degree of freedom, the contribution can vanish in certain parts of the BZ, with important consequences for the pair-breaking capabilities of a single impurity.

The symmetry-enforced zeros of the Bloch eigenstates can be obtained from the little group representations~\cite{zak:1981,bradlyn:2017,cano:2018,cano:2021}. If the Hamiltonian $\mathcal{H}$ is invariant under the space group $G$, we can label the eigenstates of the Hamiltonian by irreducible representations (irreps) of the little group at $\bm{k}$. In the case of only one-dimensional irreps, this implies that for each element $g_{\bm{k}}$ in the little group $G_{\bm{k}}$ of $G$ we have
\begin{equation}
    \rho^{\bm{k}}(g_{\bm{k}}) \bm{u}^{\kappa}(\bm{k}) = \lambda^{\kappa}(g_{\bm{k}}) \bm{u}^{\kappa}(\bm{k})\,, \label{eq:band_rep_eigenvalue}
\end{equation}
where $\rho^{\bm{k}}(g_{\bm{k}})$ is a matrix representation of $g_{\bm{k}}$ and $\kappa$ labels the irrep of the band at $\bm{k}$ with character $\lambda^{\kappa}(g_{\bm{k}})$.
The implications of Eq.~\eqref{eq:band_rep_eigenvalue} can be illustrated as follows. We consider the Lieb lattice (space group P4mm, \#99), as shown in the inset of Fig.~\ref{fig:Figure2}(a), and assume that each site hosts only a trivial orbital, that is, one that respects all symmetries of the site. In what follows, we will assume that spin-orbit coupling is negligible, allowing us to ignore spin. As an example, we focus on the high-symmetry point Y, which is on the line connecting the M and X points, as seen in Fig.~\ref{fig:Figure2}(b)~\footnote{We follow the notation of the Bilbao Crystallographic Server}. In units of the reciprocal lattice vectors, defined as $\bm{g}_1=(2\pi,0)$ and $\bm{g}_2 = (0,2\pi)$, Y is given by $\mathrm{Y}=(v,\tfrac{1}{2})$ with $0<|v|<\tfrac{1}{2}$.
The little group at Y consists of the identity and a mirror operation, $m_{010}$, which denotes a mirror plane that bisects the A sites of Fig.~\ref{fig:Figure2}(a)~\footnote{One can equivalently choose the mirror plane bisecting the B and C sites. In this case, the designation of the $Y_{1}$ and $Y_{2}$ irreps flips when compared to the mirror plane bisecting the A site. As a result, there are two $Y_{1}$ irreps and one $Y_{2}$ irrep in this case, and the symmetry-enforced zeros of Eq.~\eqref{eq:symmetry_enforced_zero} hold with $Y_{1} \leftrightarrow Y_{2}$.}. The sites that are shifted by the mirror operation acquire phase factors and the matrix representation of the band representation of $m_{010}$ at Y is
\begin{equation}
    \rho^{\mathrm{Y}}(m_{010}) = \begin{pmatrix}
        1 & 0 & 0 \\
        0 & -1 & 0 \\
        0 & 0 & -1
    \end{pmatrix}\,.
\end{equation}
It follows that there are two possible irreps at Y, namely $Y_1$ and $Y_2$, where $Y_1$ is even under the mirror while $Y_2$ is odd. Thus, $\lambda^{Y_1}(m_{010})=+1$ and $\lambda^{Y_2}(m_{010})=-1$ in Eq.~\eqref{eq:band_rep_eigenvalue}. Moreover, one of the Bloch states transforms as $Y_{1}$ whereas the other two transform as $Y_2$. Combined with Eq.~\eqref{eq:band_rep_eigenvalue}, this forces specific components of the Bloch states to vanish. At Y, these are
\begin{equation}
    u^{Y_2}_{\mathrm{A}}(\mathrm{Y}) = u^{Y_1}_{\mathrm{B}}(\mathrm{Y}) = u^{Y_1}_{\mathrm{C}}(\mathrm{Y}) = 0\,. \label{eq:symmetry_enforced_zero}
\end{equation}

A similar procedure can be applied to the other points in the BZ hosting one-dimensional irreps and restricts the $\bm{k}$-dependence of the Bloch states. The result of this analysis is illustrated by the pink lines in Fig.~\ref{fig:Figure2}(b), which denote the points where symmetry enforces the vanishing of specific components of the Bloch states.
In Fig.~\ref{fig:Figure2}(a) we show the electronic structure on the Lieb lattice; the shaded lines, also shown in panel (b), are corresponding results including a next-nearest neighbor hopping parameter $t'$.
Here, the A and C sites correspond to the 2c Wyckoff position of the P4mm space group, while B corresponds to the 1b Wyckoff position.
Figure~\ref{fig:Figure2}(b) shows the Bloch weights $|u_{\alpha}^{n}(\bm{k})|^2$ for each band and sublattice site. The contours indicate $|u_{\alpha}^{n}(\bm{k})|^2=0.1$ for the values of $t'$ indicated in Fig.~\ref{fig:Figure2}(a). The results agree with the symmetry analysis above and shows that $t'$ controls the degree of anisotropy in specific components of the Bloch weights.

In the Supplementary Material (SM)~\cite{SM}, we generalize the above considerations to other lattices and to non-trivial orbitals.
We emphasize that the zero constraints derived for the Bloch weights depend solely on the orbital content and their Wyckoff position, rather than on specific Hamiltonian parameters, highlighting the robustness of certain features of the observed sublattice weight distribution to the inclusion of, for example, longer-range hoppings.

\paragraph{Superconductivity and $T_{\mathrm{c}}$ suppression.}
The symmetry-enforced nodes and the associated suppressed weight of the Bloch functions have direct consequences for the robustness of superconductivity. To demonstrate this property, we have calculated the disorder-averaged suppression of the order parameter and the critical temperature $T_{\mathrm{c}}$ within the AG framework~\cite{abrikosov:1960}. Although this method neglects spatial inhomogeneity and other feedback effects~\cite{romer:2018,gastiasoro:2013,gastiasoro:2016,gastiasoro:2018}, it provides a simple method to compare the robustness of various superconducting pairing states to disorder. In AG-theory, the full Green function is
\begin{align}
\label{eq:fullGreen}
    \widehat{G}(\bm{k},i\omega_n)^{-1} = \widehat{G}^{(0)}(\bm{k},i\omega_n)^{-1} - \widehat{\Sigma}(
    %\bm{k},
    i\omega_n),
\end{align}
where $\widehat{G}^{(0)}(\bm{k},i\omega_n)$ is the Green function of the superconductor and $\widehat{\Sigma}(
%\bm{k},
i\omega_n)$ denotes the local self-energy. In the Born approximation, $\widehat{\Sigma}(
%\bm{k},
i\omega_n)$ is
\begin{align}\label{eq:selfenergy}
    \widehat{\Sigma}(i\omega_n) = \frac{n_{\mathrm{imp}}V^{2}}{N}\sum_{\alpha \bm{k}} \widehat{h}_{\mathrm{imp}}^{\alpha} \widehat{G}^{(0)}(\bm{k},i\omega_n) \widehat{h}_{\mathrm{imp}}^{\alpha},
\end{align}
where $n_{\mathrm{imp}}$ is the impurity density and $\widehat{h}_{\mathrm{imp}}^{\alpha}$ refers to the local non-magnetic impurity potential without the impurity strength prefactor, that is $\widehat{H}_{\mathrm{imp}}^{\alpha}=V\widehat{h}_{\mathrm{imp}}^{\alpha}$.
The superconducting order parameter is given by
\begin{align}
\label{eq:orderparam}
    \Delta_{\alpha\beta}(\bm{k}) = \frac{T V_{\mathrm{SC}}}{N} \left(f_{\bm{k}}^{\beta\alpha}\right)^{\ast} \sum_{\substack{ \bm{k}' \omega_n \\ \gamma\delta}} f^{\gamma\delta}_{\bm{k}'} G_{\gamma\bar\delta}(\bm{k}',i\omega_n)\,.
\end{align}
The above expression corresponds to an assumed pairing interaction of the form $V^{\alpha\beta\gamma\delta}_{\bm{k},\bm{k}'} = -V_{\mathrm{SC}} \left(f_{\bm{k}}^{\beta\alpha}\right)^{\ast}f^{\gamma\delta}_{\bm{k'}}$, where $V_{\mathrm{SC}}$ is the pairing strength, and $f^{\alpha\beta}_{\bm{k}}$ is a form factor transforming as the irrep of interest~\cite{holbaek:2023, SM}. The two inter-dependent equations, (\ref{eq:fullGreen}) and (\ref{eq:orderparam}),  are solved selfconsistently to obtain the order parameter $\Delta(\bm{k})$ as a function of temperature $T$.

\begin{figure*}[t]
\centering
\includegraphics[width=\linewidth]{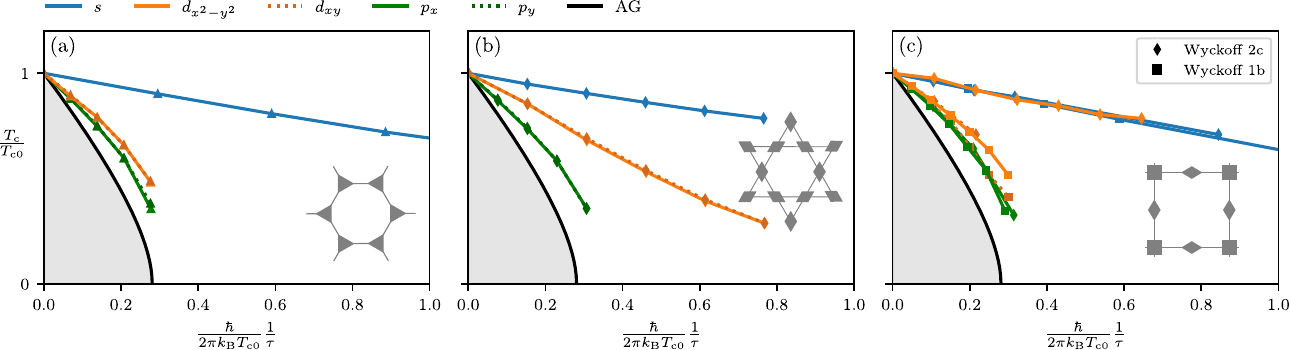}
\caption{\label{fig:Tcsuppression}\textbf{Calculated $T_{\mathrm{c}}$-suppression for the honeycomb, kagome, and Lieb lattices.} The black curve shows the solution of the AG equation $\ln(\frac{T_{\mathrm{c}}}{T_{\mathrm{c}0}}) = \psi(\frac{1}{2}) - \psi(\frac{1}{2} + \frac{\hbar}{4\pi k_{\mathrm{B}}T_{\mathrm{c}}}\frac{1}{\tau})$~\cite{mineev:1999}, with $\psi$ the digamma function. Unconventional superconductivity on the honeycomb lattice (a) exhibits the standard behavior of $T_{\mathrm{c}}$ versus scattering rate. In contrast, some of the unconventional order parameters on the kagome (b) and Lieb (c) lattices are remarkably robust to disorder, as seen by the orange curves with diamond markers in panels (b) and (c). The parameters used to generate each figure are listed in the SM~\cite{SM}.}
\end{figure*}

In Figure~\ref{fig:Tcsuppression}, we plot $T_{\mathrm{c}}$ as a function of the scattering rate $\tau^{-1}$ for a range of symmetry-distinct superconducting order parameters (colored lines) for the cases of the simple honeycomb (a), kagome (b), and Lieb (c) lattices. The scattering rate from impurities on sublattice $\alpha$, calculated from Fermi's golden rule generalized to the multi-sublattice case, is $\frac{1}{\tau^{\alpha}} = \frac{2\pi}{\hbar} n_{\mathrm{imp}}^{\alpha}V^{2} \braket{|u_{\alpha}^{n^{\ast}}(\bm{k})|^{2}}_{\text{FS}} \rho_{\alpha}(0)$, where $\braket{\ldots}_{\mathrm{FS}}$ denotes a Fermi surface average, and $\rho_{\alpha}(0)$ is the sublattice-projected density of states (DOS), calculated by averaging the sublattice-resolved DOS over an energy range corresponding to the magnitude of the gap~\cite{SM}.
For the honeycomb and kagome lattices, the sublattice sites correspond to identical Wyckoff position and impurities are distributed on each site with equal probability. For the Lieb lattice on the other hand, the sublattice sites A and C are symmetry distinct from site B, and we calculate
separate $T_{\mathrm{c}}$ suppression curves for impurities on Wyckoff position 2c (sites A and C) and for impurities on position 1b (site B). The total scattering rate is obtained from summing the scattering rates over sublattices.

As seen from Fig.~\ref{fig:Tcsuppression}(a), superconductivity on the simple honeycomb lattice exhibits the expected $T_{\mathrm{c}}$-suppression behavior, with the sign-changing orders closely following the AG curve. On the kagome lattice, only the $p$-wave order follows the AG curve, while the fully-compensated components of the $E_{2}$ ($d$-wave) order exhibits a remarkable robustness to local disorder, as we highlighted in a previous work~\cite{holbaek:2023}. Figure~\ref{fig:Tcsuppression}(c) shows the $T_{\mathrm{c}}$ suppression curves for the Lieb lattice with impurities located only on Wyckoff position 2c (diamond markers) or 1b (square markers). The $B_{1}$ ($d_{x^{2}-y^{2}}$-wave) order parameter is remarkably robust when impurities occupy the 2c position, whereas the $T_{\mathrm{c}}$ of the other unconventional superconducting orders is strongly suppressed. By contrast, impurities on the 1b position lead to the expected pair-breaking behavior for all sign-changing superconducting orders. In general, we find that order parameters transforming trivially under the symmetry elements of the site-symmetry group of the impurity-hosting sites can exhibit robustness to point-like disorder. Robust unconventional superconductivity exists also for the case of nontrivial orbitals on the Lieb lattice, as discussed in the SM~\cite{SM}.

\begin{figure}[t]
\centering
\includegraphics[width=\linewidth]{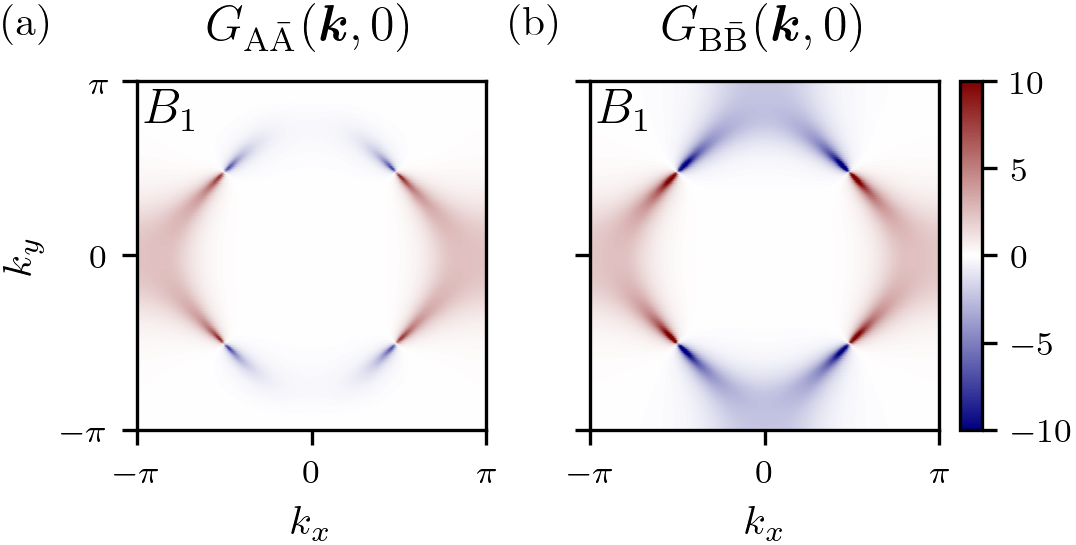}
\caption{\label{fig:Figure4}\textbf{Origin of robust superconductivity.} The approximate anomalous Green function at sublattice A (a) and B (b) of the Lieb lattice with on-site $d_{x^{2}-y^{2}}$-wave superconductivity. On the 2c Wyckoff position (sublattices A and C), the anomalous Green function for a $d_{x^{2}-y^{2}}$-wave order averages to a finite value.}
\end{figure}

The weak $T_{\mathrm{c}}$ suppression for $d$-wave orders on the kagome and Lieb lattices can be understood from the non-uniform Bloch sublattice weights. As seen in Eq.~\eqref{eq:selfenergy}, the local impurity potential projects the self-energy onto sublattice $\alpha$, leaving nonzero components only in the diagonal entries of the normal and anomalous Nambu blocks given by $\Sigma_{\alpha\alpha}$ ($\Sigma_{\bar{\alpha}\bar{\alpha}}$) in the normal electron (hole) part and $\Sigma_{\alpha\bar{\alpha}}$, $\Sigma_{\bar{\alpha}\alpha}$ in the anomalous part. In particular, $\Sigma_{\alpha\bar{\alpha}}(i\omega_{n}) \propto \sum_{\bm{k}}G_{\alpha\bar{\alpha}}^{(0)}(\bm{k},i\omega_{n})$. Approximating the Green function in the band basis by $G_{n^{\ast}\bar{n}^{\ast}}^{(0)}(\bm{k}, i\omega_{n}) \approx \frac{\Delta_{n^{\ast}}(\bm{k})}{(i\omega_{n})^{2} - \xi_{n^{\ast}}^{2}(\bm{k}) - |\Delta_{n^{\ast}}(\bm{k})|^{2}}$, where $n^{\ast}$ denotes the band crossing the Fermi energy, and all other components zero, and transforming from band to sublattice basis one finds~\cite{holbaek:2023}
\begin{equation}
    G_{\alpha\bar{\alpha}}^{(0)}(\bm{k},i\omega_{n}) \approx \frac{\Delta_{n^{\ast}}(\bm{k})|u_{\alpha}^{n^{\ast}}(\bm{k})|^{2}}{(i\omega_{n})^{2} - \xi_{n^{\ast}}^{2}(\bm{k}) - |\Delta_{n^{\ast}}(\bm{k})|^{2}}.
\end{equation}
Thus, momentum-independent Bloch sublattice weights yield a vanishing numerator of the self-energy for an unconventional (sign-changing) order parameter. The simple honeycomb lattice, with only two sites per unit cell, provides a representative example with constant sublattice weights $|u_{\alpha}^{n}(\bm{k})|^2 = 1/2$, excluding any extended regions with vanishing Bloch weight, and the $T_{\mathrm{c}}$-suppression curve therefore closely follows the AG result. In contrast, a nontrivial momentum dependence of the Bloch weights can produce finite anomalous contributions even for a sign-changing $d$-wave order, which influences the $T_{\mathrm{c}}$ suppression. Figure~\ref{fig:Figure4} illustrates this using the anomalous Green function for a $B_{1}$ ($d_{x^{2}-y^{2}}$) order parameter on the Lieb lattice. Due to the non-uniform Bloch weight on sublattice A, its anomalous component averages to a finite value, similar to what happens for an $s$-wave superconductor, which results in a significantly weaker $T_{\mathrm{c}}$ suppression for impurities on that site. On the other hand, the Bloch weight on sublattice B preserves the full symmetry of the lattice and therefore yields a vanishing component of the anomalous Green function, leading to the usual AG behavior for impurities on that site. This mechanism for robust unconventional superconductivity is distinct from the robustness discussed in other theoretical works~\cite{michaeli:2012,timmons:2020,andersen:2020,cavanagh:2020,zinkl:2022}.

The protection of unconventional superconductivity to disorder may be realized in kagome superconductors, including the $A$V$_3$Sb$_5$ ($A$: K, Rb, Cs) compounds. At present, however, the symmetry properties of the superconducting ground state of these materials remain controversial~\cite{wilson:2024,romer:2022,he:2025}. Any kagome material with $E_2$ ($d$-wave) pairing should exhibit very robust superconductivity to point-like disorder. In terms of superconductivity on the Lieb lattice, the cuprates are prime examples. Our study reveals that the $d_{x^2-y^2}$ pairing in the cuprates should be very robust to O defect disorder. This is complicated however, by the fact that superconductivity in the cuprates arises from doping a charge transfer insulator with the Cu $d$-orbitals as the important states. More direct manifestations of the current robustness should arise for superconductors in inverse Lieb materials~\cite{chang:2025,jiang:2025,wei:2025}, where the important orbitals for superconductivity originate from atoms at the 2c Wyckoff site. Unconventional order parameters in such materials have recently become of interest due to their relation to altermagnetism~\cite{wu:2025}. We are not aware of realizations of unconventional superconductivity in such compounds, but once found, they provide promising platforms for direct tests of the scenario presented in this work. Alternatively, it might be possible to verify it in engineered structures such as imperfect artificial lattices or patterned structures, or proximitized systems of unconventional superconductors with Lieb or kagome metals containing controllable amounts of disorder.

\paragraph{Conclusions.}
In summary, we have elucidated the general circumstances under which unconventional superconductivity can be protected from disorder, and illustrated this through very weak $T_{\mathrm{c}}$ suppression for the kagome and Lieb lattices. These are simple two-dimensional cases where nontrivial sublattice-dependent momentum distributions of the Bloch states restore the robustness of unconventional superconductivity to disorder. For the same reason, there are no in-gap impurity bound states. We stress that many other physical observables may well also exhibit non-standard behavior due to this effect. For the kagome lattice, for example, the spin-lattice relaxation rate features a pronounced Hebel-Slichter peak even for a fully compensated $d$-wave gap symmetry~\cite{dai:2024}. However, this is merely one example, and future explorations of unconventional superconductivity in space groups featuring multiple low-energy orbitals per unit cell at low-symmetry Wyckoff positions may uncover more surprises.

\begin{acknowledgments}
\paragraph{Acknowledgments.} 
We acknowledge Luca Buiarelli, Mark H. Fischer, Titus Neupert, and Martina O. Soldini for helpful discussions. M.H.C acknowledges support by ERC grant project 101164202 -- SuperSOC. Funded by the European Union. Views and opinions expressed are however those of the authors only and do not necessarily reflect those of the European Union or the European Research Council Executive Agency. Neither the European Union nor the granting authority can be held responsible for them. A.~K. acknowledges support by the Danish National Committee for Research Infrastructure (NUFI) through the ESS-Lighthouse Q-MAT. B.M.A. acknowledges support from the Independent Research Fund Denmark Grant No. 5241-00007B.
\end{acknowledgments}

\bibliography{bibliography}

\end{document}